# INTRODUCTION AND STATUS OF FERMILAB'S ACORN PROJECT*

D. Finstrom[†], E. Gottschalk, Fermilab, Batavia, USA

*Abstract*

Modernizing the Fermilab accelerator control system is essential to future operations of the laboratory's accelerator complex. The existing control system has evolved over four decades and uses hardware that is no longer available and software that uses obsolete frameworks. The Accelerator Controls Operations Research Network (ACORN) Project will modernize the control system and replace end-of-life power supplies to enable future accelerator complex operations with megawatt particle beams. An overview of the ACORN Project and a summary of recent research and development activities will be presented.

## INTRODUCTION

The Fermilab Accelerator Complex is the largest accelerator complex in the United States and the second largest in the world. It currently operates with a single control system, ACNET [1], that contains hardware and software for controlling ten miles of accelerator components and beam transfer lines. The control system was originally developed for the start of colliding-beam operations in 1983. The control system initiates particle beam production, controls beam energy and intensity, transports particle beams to research facilities, measures beam parameters, and monitors beam transport through the accelerator complex to ensure safe, reliable, and effective operations. There are approximately 200,000 devices with 350,000 attributes and several million lines of software code in the existing system. Despite operating at peak performance, in 2018, an advisory committee formed to evaluate facility operational risks identified the need to invest in the accelerator control system and called out major issues, including a large amount of old hardware and software and an aging and declining in strength workforce with no software development related hires for 18 years. The Accelerator Controls Operations Research Network (ACORN) Project addresses these concerns and will replace control system hardware that is no longer available and software that is no longer maintainable. ACORN is a U.S. Department of Energy (DOE) Project for Acquisition of Capital Assets (O413.3B) to modernize the accelerator control system and replace end-of-life power supplies to enable future operations of the accelerator complex with megawatt particle beams. The total project cost range is $100 - $142 million dollars USD. The modernized control system will integrate with two projects, the Long Baseline Neutrino Facility/Deep Underground Neutrino Experiment (LBNF/DUNE) and the Proton Improvement Plan II (PIP-II). ACORN received Critical Decision 0 (CD-0) milestone from DOE, which signifies approval of the project's mission need, on August 28, 2020. ACORN is working towards Critical Decision 1 (CD-1) milestone, which signifies approval of the alternative selection and cost range and is targeted to occur in Q3 FY24. As required by CD-1, the project is evaluating design alternatives and, consequently, has not settled on a technical design. Efforts are focused on building the project team, interviewing stakeholders, gathering requirements, and research and development efforts aimed at determining project cost and schedule estimates. Critical Decision 4 (CD-4) milestone, which signifies project completion, is projected to occur in the 2028 – 2030 time frame. The project Work Breakdown Structure consists of five main sections: Project Management, Accelerator Power Systems, Data Acquisition and Control, Control System Infrastructure, and Control System Applications.

## ACCELERATOR POWER SYSTEMS

The scope of work for the Accelerator Power Systems (APS) section includes the design, prototype, procurement, installation, and testing of power supplies and regulation systems for the Fermilab Accelerator Complex. Two types of power supplies are being considered: switch-mode power supplies and thyristor power supplies. The existing power supplies are almost all original and date back to the 1970s and 1980s. The power supplies were originally bid and built to specification and have a straightforward design and rugged construction. Fermilab has continued to buy power supplies using a combination of procurement strategies consisting of build-to-specification procurements for power components and build-to-print procurements for control components. Following the guidance of subject matter experts and our experience at Fermilab, we plan to separate the procurement of power conversion components from components needed for power regulation and control. The power conversion components are available from commercial vendors, and competition among vendors keeps costs reasonable. Power regulation components are highly specialized and can be applied to different vendors. This helps establish commonality among power supply systems and allows the replacement of power converters when requirements change.

Hiring expertise in accelerator power systems has been challenging, and consequently, progress in this section has been constrained. ACORN is working with Fermilab's Electrical Engineering Support department to understand the hazards associated with the installation, testing, and maintenance of power supplies and working with the ACORN Environment, Safety, and Health liaison to begin writing the preliminary hazard analysis report for CD-1. The project team is compiling an inventory of power

---

supplies, creating diagrams for each configuration, and continuing to build out the team. To address staffing, ACORN is currently working on establishing an agreement with the Electrical Power Conversion Group at Oak Ridge National Laboratory to collaborate on both project management and electrical engineering work for ACORN's accelerator power systems.

## DATA ACQUISITION AND CONTROL

The scope of work for Data Acquisition and Control (DAC) includes field hardware systems and infrastructure. The existing field hardware includes general purpose digital and analog input/output channels that are primarily CAMAC-based hardware, as well as front-end systems that connect the hardware to the greater control system. The field system infrastructure includes links that provide signals ranging from timing to machine protection to communication with the greater control system, networking, and hardware infrastructure, such as the backplane design for crate systems.

Work in this area has focused on building the team, conducting interviews to gather requirements and identify use cases, developing an equipment inventory, and creating a general purpose field system concept. The leadership team needed for CD-1 was fully established in late September 2023. Extensive stakeholder interviews were conducted earlier in the year to identify use cases and technical requirements that were used to synthesize the specifications for new hardware systems and components. An inventory application was developed to explore potential replacement options, and a general-purpose field system concept was developed.

The DAC group is actively pursuing research and development (R&D) to explore MicroTCA-based solutions to replace existing modular, general purpose chassis systems. MicroTCA (MTCA) allows a high density of devices with rear-transition modules using existing cables. The systems allow monitoring and control of remote power and cooling and have hot-swappable modules. These ideas are based on existing Fermilab experience with PIP-II instrumentation work with Vadatech. The MTCA subsystem can incorporate an FPGA and ARM processor structure on each card to standardize the design. This will provide the ability to run edge applications or Artificial Intelligence and Machine Learning (AI/ML) applications as needed. The MTCA crates will either utilize Ethernet or PCEIe across the backplane between cards. PCIe will be utilized if there is a need for fast communication between cards within a crate.

R&D work is continuing along several fronts. The DAC group is exploring new Programmable Logic Controllers (PLCs) and Human-Machine Interfaces (HMIs) and building a general-purpose test stand for developing and testing new installations that incorporate Experimental Physics and Industrial Control System (EPICS) [2] capabilities. The group is testing publish/subscribe protocols like MQTT [3] and OPC-UA [4], which are built into many new PLCs. An event-driven framework, Node-RED [5], is being investigated to serve as a front-end technology. Node-RED is built on the open-source community supported software for the Industrial Internet of Things over MQTT, OPC-UA, Modbus TCP, etc. This would enable the straightforward development of test stands for testing controls functionality and front end development outside of the production control system. There is also development work with motion control testing Step Pak Ethernet-based motion controllers and creating a motion-control test stand. Future work includes exploring a collaboration with the SLAC National Accelerator Laboratory on a common FPGA firmware architecture. The group will continue to gather technical requirements, iterate on the requirements with stakeholders, and develop datasheets for a set of concept hardware modules to replace existing obsolete field hardware.

## CONTROL SYSTEM INFRASTRUCTURE

The scope of work for Control System Infrastructure (CSI) includes central services, networking, and integration and systems testing. The CSI group will deliver modernized central services to replace obsolete functionality and implement new functionality as defined by stakeholder requirements and an analysis of capability gaps. It will define the topology for wired networking, including virtual local area networks, access control lists, and network router and switch configuration. It will also employ modern processes and deliver infrastructure to streamline the integration of control system components, including an automated pipeline to build, test, and deploy code to the production environment using a Continuous Integration/Continuous Deployment methodology and provide test system infrastructure.

R&D work includes the implementation of a computing cluster for rapid prototyping using Kubernetes [6]. Kubernetes is a portable, extensible, open source platform for managing containerized workloads and services that facilitates both declarative configuration and automation. Recent work on the cluster was done to provide support for EPICS control system research and testing. Installations of Prometheus [7] and Grafana [8] were added for monitoring and observability. A data lake is being investigated for archiving data. The data lake is envisioned as a system with services that manage data ingest, data transformations, data storage, and data queries. Also included are computing and disk resources, user-supplied code for cataloguing and data transforms, and interfaces for writing and accessing data. The system is envisioned as a data repository for AI/ML models and training data. Existing ACNET protocols and EPICS protocols will need to be supported to facilitate the transition from ACNET to EPICS. Efforts will be made to consolidate exposed protocols as much as possible and handle protocol translations internally. Data acquisition services will be handled by a Data Pool Manager (DPM). DPM will abstract protocols from end-users to support EPICS integration. Data stress tests have been conducted, and work with the Control System Applications group is ongoing to define processes for building and deploying applications. After prototyping and evaluating tools, GitHub Actions was selected for testing and compiling software. Several deployment tools are being evaluated. A Role Based Access Control (RBAC) tool, Keycloak [9], has been selected and installed, and work is ongoing to explore authorization and implicit workflows.

## CONTROL SYSTEM APPLICATIONS

Control System Applications (CSA) encompasses the following deliverables: control system application framework, user applications, and user interface. The user interface deliverable includes a style guide that is being developed by human factors experts from Idaho National Laboratory (INL).

R&D efforts on control system applications are driven by user requirements, with priority given to requirements that have been identified by control room operators. One example of user requirements driving R&D evaluations is the need to explore the use of web applications for accelerator operations. Three browser application frameworks (React [10], Flutter [11], and Fresh [12]) were evaluated. This effort resulted in the selection of Flutter as the framework that is preferred for R&D purposes.

The CSA group is collaborating with developers working on the design and development of a PIP-II parameter display application. When completed, the effort will enable fast feedback on design concepts that will be used to develop cost estimates and help define interfaces for integrating PIP-II and ACORN applications. Approximately 800 legacy applications, totalling several million lines of code, are under consideration for conversion. A detailed analysis of these applications, along with stakeholder requirements, was used to determine critical applications, applications to be retired, and applications to be consolidated. Currently, an evaluation of application complexity is underway. The complexity of the conversion process requires accurate labor estimates to determine the amount of effort needed to modernize applications.

The CSA group is also working with the INL human factors experts on a style guide and application templates that include updates based on feedback from the PIP-II collaboration. This will provide graphical user interface implementers with clear guidance for a consistent and accessible user experience.

## ENABLING CAPABILITIES

The ACORN project will enable new capabilities for the accelerator control system and the requirements are driven by future needs for accelerator operations. For example, AI/ML capabilities are being explored at Fermilab to optimize future accelerator operations, predict failure modes, and provide new simulation capabilities. These AI/ML R&D efforts have provided insights that demonstrate the need to include support for AI/ML capabilities as part of a modernized accelerator control system. Use cases have been identified and stakeholder requirements will be used to develop the design of the accelerator control system. The most significant need that has been identified is the need to support Machine Learning Operations (MLOps), which will help standardize and streamline the end-to-end machine learning lifecycle from development and training to deployment, monitoring, and maintenance.

## CONCLUSION

With recent efforts focused on completing the project team in preparation for future project reviews, the rate of progress on the ACORN project has greatly increased over the last few months. The project team's primary focus is on preparations for the CD-1 milestone in 2024. Recent R&D efforts have focused on establishing cost and schedule estimates that will be needed for CD-1 and on evaluating accelerator control system design alternatives required for CD-1.

## ACKNOWLEDGEMENT

The authors would like to thank ACORN team members Maria Acosta Flechas, Lila Anderson, Jennifer Case, Jonathan Eisch, Beau Harrison, Chris Roehrig, and Anthony Tiradani for their contributions.

## REFERENCES

[1] K. Cahill *et al*., "The Fermilab Accelerator Control System," *ICFA Beam Dyn. Newslett.,* vol.47, pp. 106-124, FERMILAB-PUB-08-605-AD, 2008.

[2] EPICS, https://epics-controls.org

[3] MQTT, https://mqtt.org

[4] OPC-UA, https://opcfoundation.org/about/opc-technologies/opc-ua

[5] Node-RED, https://nodered.org

[6] Kubernetes, https://kubernetes.io

[7] Prometheus, https://prometheus.io

[8] Grafana, https://grafana.com

[9] Keycloak, https://keycloak.org

[10] React, https://react.dev

[11] Flutter, https://flutter.dev

[12] Fresh, https://fresh.deno.dev